\documentclass{nic-series}
\begin{document} 
\title{Phase separation in peptide aggregation processes --\\ Multicanonical study of a mesoscopic model}
\author{Christoph Junghans \inst{1,2} \and Michael Bachmann \inst{1,3} \and Wolfhard Janke \inst{1}}
\institute{Institut f\"ur Theoretische Physik and Centre for Theoretical Sciences (NTZ), \\
        Universit\"at Leipzig, Postfach 100\ 920, D-04009 Leipzig, Germany\\
        \email{janke@itp.uni-leipzig.de}
	\and
	Max-Planck-Institut f\"ur Polymerforschung, Ackermannweg 10,
	D-55128 Mainz, Germany\\
	\email{junghans@mpip-mainz.mpg.de}
	\and
	Computational Biology \& Biological Physics Group, Department of Theoretical Physics,\\
	Lunds Universitet, S\"olvegatan 14A, SE-223 62 Lund, Sweden\\
	\email{bachmann@thep.lu.se}
}
\maketitle
\begin{abstracts}
We have performed multicanonical computer simulations of a small system of short protein-like 
heteropolymers and found that their aggregation transition possesses similarities to first-order phase 
separation processes. Not being a phase transition in the thermodynamic sense, the observed folding-binding
behavior exhibits fascinating features leading to the conclusion that the temperature is no suitable
control parameter in the transition region. More formally, for such small systems the microcanonical 
interpretation is more favorable than the typically used canonical picture.
\end{abstracts}
\section{Introduction}
Folding-binding and docking processes between proteins are significant for catalysis, transport, 
and cell stabilization in biological systems. Also, gene replication and expression are 
impossible without defined binding mechanisms of molecules. However, the mutual influence of proteins 
on each other can also result in refolding of proteins (which often leads to the loss of their functionality 
and thus biological activity) or cluster formation. In the latter case, proteins self-assemble and 
form aggregates. The effects of plaque can be disastrous and cause heavy diseases:
First, the assembled proteins lose their individual functionality and second, in the passive case,
the aggregates hinder transport and signal exchange processes which are significant for the life of cells. 
In an active process, specific aggregates might be able to bind to cell membranes and
to change the membrane morphology, e.g., by forming pores. In the amyloid hypothesis for the onset
of Alzheimer's disease, for example, aggregates of A$\beta$ proteins are believed to form pores in membranes
of neuron cells, thus opening ion channels for neurotoxic calcium.~\cite{pores}

We focus here on thermodynamic properties of the aggregation transition of small peptides. 
For this purpose, a 
simple hydrophobic-polar aggregation model is introduced and employed in a multicanonical study
of a few short heteropolymers.~\cite{jbj1,jbj2}
\section{Aggregation model}
For the aggregation study, we extend the AB model~\cite{still1} by an additional interchain
interaction between the $M$ heteropolymers. As in the single-chain model, which has proven 
to be quite useful in qualitative studies of tertiary folding behavior,~\cite{ssbj}
only two types of amino acids are considered: hydrophobic residues (A) which avoid contact with the polar
environment and polar residues (B) being favorably attracted by the solvent. The single-chain
energy of the $\mu$th heteropolymer ($\mu=1,\ldots,M$) composed of $N_\mu$ monomers is given by~\cite{still1}
\begin{equation}
\label{eq:abmod}
E_{\rm AB}^{(\mu)}=\frac{1}{4}\sum\limits_{i_\mu}(1-\cos \vartheta_{i_\mu})+%
\!\!\sum\limits_{j_\mu>i_\mu+1}\Phi(r_{i_\mu j_\mu};\sigma_{i_\mu},\sigma_{j_\mu}),
\end{equation}
where $0\le \vartheta_{i_\mu}\le \pi$ denotes the virtual bending angle between the monomers 
$i_\mu$, $i_\mu+1$, and $i_\mu+2$. Not discriminating nonbonded interactions
between monomers of the same or different polymers, our aggregation model 
reads~\cite{jbj1}
\begin{equation}
\label{eq:aggmod}
E=\sum\limits_{\mu} E_{\rm AB}^{(\mu)}+\sum\limits_{\mu<\nu} 
\sum_{i_\mu,j_\nu}\Phi(r_{i_\mu j_\nu};\sigma_{i_\mu},\sigma_{j_\nu}),
\end{equation}
where $\mu,\nu$ label the $M$ polymers interacting with each other, and 
$i_\mu, j_\nu$ index the monomers of the respective $\mu$th and $\nu$th polymer.
The nonbonded interresidue pair potential 
$\Phi(r_{i_\mu j_\nu};\sigma_{i_\mu},\sigma_{j_\nu})=
4[r_{i_\mu j_\nu}^{-12}-C(\sigma_{i_\mu},\sigma_{j_\nu})r_{i_\mu j_\nu}^{-6}]$
depends on distance $r_{i_\mu j_\nu}$ and residue type
$\sigma_{i_\mu}=A,B$. The long-range behavior is attractive for 
like pairs of residues [$C(A,A)=1$, $C(B,B)=0.5$] and repulsive otherwise [$C(A,B)=C(B,A)=-0.5$]. 
The length of all virtual peptide bonds is unity. In this short note, we focus on a system of two identical
chains with the Fibonacci sequence $AB_2AB_2ABAB_2AB$, where the single-chain properties are
known.~\cite{baj1} Our primary interest is devoted to the phase behavior of the system and 
for this purpose, the density of states $g(E)$ is a suitable quantity that we
obtained by means of multicanonical computer simulations.~\cite{muca} 
\section{Microcanonical vs.\ canonical view}
\begin{figure}
\centerline{\epsfxsize=8.5cm \epsfbox{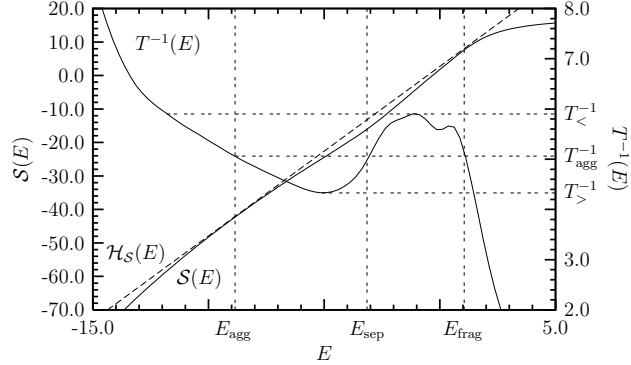}}
\caption{\label{fig:micro} Microcanonical Hertz entropy ${\cal S}(E)$, concave Gibbs hull 
${\cal H}_{\cal S}(E)$, and inverse caloric temperature $T^{-1}(E)$ as functions of energy.
The phase separation regime is bounded by $E_{\rm agg}$ and $E_{\rm frag}$; the 
temperature region, where temperature is no suitable external control parameter
and the canonical interpretation breaks down, ranges from $T^{-1}_<$ to $T^{-1}_>$.
The slope of the Gibbs hull defines the aggregation temperature, 
$T_{\rm agg}^{-1}=\partial{\cal H}_{\cal S}(E)/\partial E={\rm const.}$}
\end{figure}

The Hertz definition of the entropy is given by ${\cal S}(E)=k_B\ln\,\Gamma(E)$, where $k_B$
is the Boltzmann constant ($k_B=1$ in our simulations) and 
$\Gamma(E)=\int_{E_{\rm min}}^E dE'\,g(E')$ (where $E_{\rm min}$ is the ground-state energy)
is the phase space volume. In Fig.~\ref{fig:micro}, ${\cal S}(E)$ is shown for our two-peptide
system. Interestingly, in the energy region between $E_{\rm agg}$ and $E_{\rm frag}$
the entropy exhibits a convex behavior, which is a strong indication for surface effects within this
small system.~\cite{gross1} Also shown in Fig.~\ref{fig:micro} is the corresponding concave hull
${\cal H}_{\cal S}(E)$, i.e., the Gibbs construction. The surface entropy, defined as 
$\Delta {\cal S}(E)={\cal H}_{\cal S}(E)-{\cal S}(E)$ is maximal at the
energy $E_{\rm sep}$. The reason for the nonvanishing surface entropy is that the transition
between the fragmented, i.e., separated chains, and the formation of a joint aggregate is
a process with phase separation which is ``delayed'' due to steric surface effects reducing the 
entropy of the total system. Since entropy reduction is only achieved by additional energy  
consumption, the surprising side effect is that in the transition regime the aggregate becomes 
colder with increasing system energy. This is verified by considering the caloric temperature
which is defined via $T^{-1}(E)=\partial{\cal S}(E)/\partial E$, also shown in Fig.~\ref{fig:micro}.
Actually, in the transition region, $T^{-1}(E)$ bends back with increasing energy. 

Consequently, there is no unique mapping between temperature and energy in the transition region
(or more precisely, within the bounds $T^{-1}_<$ and $T^{-1}_>$ indicated in Fig.~\ref{fig:micro}).
Thus the temperature should not be considered as a suitable external control parameter. From
a statistical point of view this means that for transitions with phase separation in small
systems a microcanonical interpretation is preferred over the typically 
used canonical formalism. Since the backbending effect in the peptide aggregation process
is a real physical effect, it should also
be accessible to experimental verification, as it has indeed already been observed, for example, in experiments of 
sodium cluster formation processes.~\cite{schmidt1}
\section*{Acknowledgments}
This work is partially supported by the DFG under Grant
No.\ JA 483/24-1/2 and the computer time Grant No.\ hlz11 of NIC, 
Forschungszentrum J\"ulich. M.B.\ thanks the DFG and Wenner-Gren Foundation
for research fellowships. We are also grateful to DAAD-STINT for a Personnel Exchange Programme 
with Sweden.

\end{document}